\documentclass[fleqn,usenatbib]{mnras}

\usepackage[T1]{fontenc}
\usepackage{ae,aecompl}

\usepackage{graphicx}	
\usepackage{amsmath}	
\usepackage{amssymb}	
\usepackage{float}
\usepackage{color}
\usepackage{scalerel}   
\usepackage{tikz}       

\newcommand\numberthis{\addtocounter{equation}{1}\tag{\theequation}}

\defcitealias{Cortese2014}{C14}
\defcitealias{Aquino-Ortiz2018}{A18}

\usetikzlibrary{svg.path}

\definecolor{orcidlogocol}{HTML}{A6CE39}
\tikzset{
  orcidlogo/.pic={
    \fill[orcidlogocol] svg{M256,128c0,70.7-57.3,128-128,128C57.3,256,0,198.7,0,128C0,57.3,57.3,0,128,0C198.7,0,256,57.3,256,128z};
    \fill[white] svg{M86.3,186.2H70.9V79.1h15.4v48.4V186.2z}
                 svg{M108.9,79.1h41.6c39.6,0,57,28.3,57,53.6c0,27.5-21.5,53.6-56.8,53.6h-41.8V79.1z M124.3,172.4h24.5c34.9,0,42.9-26.5,42.9-39.7c0-21.5-13.7-39.7-43.7-39.7h-23.7V172.4z}
                 svg{M88.7,56.8c0,5.5-4.5,10.1-10.1,10.1c-5.6,0-10.1-4.6-10.1-10.1c0-5.6,4.5-10.1,10.1-10.1C84.2,46.7,88.7,51.3,88.7,56.8z};
  }
}

\newcommand\orcidicon[1]{\href{https://orcid.org/#1}{\mbox{\scalerel*{
\begin{tikzpicture}[yscale=-1,transform shape]
\pic{orcidlogo};
\end{tikzpicture}
}{|}}}}

\title[SAMI mass-kinematics scaling relations]{The SAMI Galaxy Survey: mass-kinematics scaling relations}

\author[D. Barat et al.]{Dilyar Barat$^{1,2}$\thanks{E-mail: Dilyar.Barat@anu.edu.au}\orcidicon{0000-0002-4972-8188},
Francesco D'Eugenio$^{3}$\orcidicon{0000-0003-2388-8172},
Matthew Colless$^{1,2}$\orcidicon{0000-0001-9552-8075},
Sarah Brough$^{4}$\orcidicon{0000-0002-9796-1363},\newauthor
Barbara Catinella$^{2,5}$\orcidicon{0000-0002-7625-562X},
Luca Cortese$^{2,5}$\orcidicon{0000-0002-7422-9823},
Scott M. Croom$^{2,6}$\orcidicon{0000-0003-2880-9197},
Anne M. Medling$^{1,7}$\thanks{Hubble Fellow},\newauthor
Sree Oh$^{1,2}$\orcidicon{0000-0002-4731-9604},
Jesse van de Sande$^{2,6}$\orcidicon{0000-0003-2552-0021},
Sarah M. Sweet$^{2,8}$\orcidicon{0000-0002-1576-2505},
Sukyoung K. Yi$^{9}$,\newauthor
Joss Bland-Hawthorn$^{2,6}$\orcidicon{0000-0001-7516-4016},
Julia Bryant$^{2,6,10}$\orcidicon{0000-0003-1627-9301},
Michael Goodwin$^{11}$,
Brent Groves$^{1,2}$,\newauthor
Jon Lawrence$^{11}$,
Matt S. Owers$^{12}$\orcidicon{0000-0002-2879-1663},
Samuel N. Richards$^{13}$\orcidicon{0000-0002-5368-0068},
Nicholas Scott$^{2,6}$\orcidicon{0000-0001-8495-8547} 
\\
$^{1}$Research School of Astronomy and Astrophysics, Australian National University, Canberra, ACT 2611, Australia\\
$^{2}$ARC Centre of Excellence for All Sky Astrophysics in 3 Dimensions (ASTRO 3D)\\
$^{3}$Sterrenkundig Observatorium, Ghent University, Krijgslaan 281, S9, 9000 Ghent, Belgium \\
$^{4}$School of Physics, University of New South Wales, NSW 2052, Australia\\
$^{5}$International Centre for Radio Astronomy Research, University of Western Australia, 35 Stirling Highway, Crawley, WA 6009,\\ ~Australia \\
$^{6}$Sydney Institute for Astronomy, School of Physics, A28, The University of Sydney, NSW 2006, Australia\\
$^{7}$Ritter Astrophysical Research Center, University of Toledo, Toledo, OH 43606, USA\\
$^{8}$Centre for Astrophysics and Supercomputing, Swinburne University of Technology, PO Box 218, Hawthorn, VIC 3122, Australia\\
$^{9}$Department of Astronomy and Yonsei University Observatory, Yonsei University, Seoul 03722, Republic of Korea\\
$^{10}$Australian Astronomical Optics, AAO-USydney, School of Physics, University of Sydney, NSW 2006, Australia\\
$^{11}$Australian Astronomical Optics, AAO-Macquarie, Faculty of Science and Engineering, Macquarie University, \\ ~~105 Delhi Rd, North Ryde, NSW 2113, Australia\\ 
$^{12}$Department of Physics and Astronomy, Macquarie University, NSW 2109, Australia\\
$^{13}$SOFIA Science Center, USRA, NASA Ames Research Center, Building N232, M/S 232-12, P.O. Box 1, Moffett Field,\\ ~~CA 94035-0001, USA
}
\date{Accepted 2019 May 22. Received 2019 May 18; in original form 2019 January 6}

\pubyear{2018}
\begin{document}
\label{firstpage}
\pagerange{\pageref{firstpage}--\pageref{lastpage}}
\maketitle

\begin{abstract}
We use data from the Sydney-AAO Multi-object Integral-field spectroscopy (SAMI) Galaxy Survey to study the dynamical scaling relation between galaxy stellar mass $M_*$ and the general kinematic parameter $S_K = \sqrt{K V_{rot}^2 + \sigma^2}$ that combines rotation velocity $V_{rot}$ and velocity dispersion $\sigma$. We show that the $\log M_* - \log S_K$ relation: 
(1)~is linear above limits set by properties of the samples and observations; 
(2)~has slightly different slope when derived from stellar or gas kinematic measurements;
(3)~applies to both early-type and late-type galaxies and has smaller scatter than either the Tully-Fisher relation ($\log M_* - \log V_{rot}$) for late types or the Faber-Jackson relation ($\log M_* - \log\sigma$) for early types; and
(4)~has scatter that is only weakly sensitive to the value of $K$, with minimum scatter for $K$ in the range 0.4 and 0.7.
We compare $S_K$ to the aperture second moment (the `aperture velocity dispersion') measured from the integrated spectrum within a 3-arcsecond radius aperture ($\sigma_{3^{\prime\prime}}$). We find that while $S_{K}$ and $\sigma_{3^{\prime\prime}}$ are in general tightly correlated, the $\log M_* - \log S_K$ relation has less scatter than the $\log M_* - \log \sigma_{3^{\prime\prime}}$ relation.
\end{abstract}


~ \\ ~ \\ ~ \\


\section{Introduction}

Galaxy scaling relations correlate observable quantities of galaxies and capture trends among physical properties. These properties can include galaxy stellar mass $(M_*)$, half-light radius $(R_e)$, rotation velocity $(V_{rot})$, velocity dispersion $(\sigma)$, luminosity $(L)$, surface brightness $(\Sigma)$ and other measurable quantities \citep{mcgaugh00, Pizagno2005, Courteau2007, Avila-Reese2008, Catinella2012}. For example, the Faber-Jackson relation \citep[FJ;][]{FaberJackson1976} connects $\sigma$ and $L$ while the Kormendy relation \citep{Kormendy1977} links $\Sigma$ and $R_{e}$. 

Galaxy scaling relations are convenient in predicting physical galaxy properties because they do not require analytic modelling of a galaxy's internal dynamics. Using scaling relations to estimate quantities such as distance and mass is efficient when the sample size is too large to obtain detailed observations or to perform individual analyses. 

Scaling relations such as the FJ and Kormendy relations have significant intrinsic scatter that impacts the precision of their predictions. Sample pruning and target selection are necessary to produce tight relations. For morphologically defined classes of galaxies, the Tully-Fisher \citep[TF;][]{tullyfisher77} relation provides a tight relation between $L$ and $V_{rot}$ for disk-dominated galaxies and the Fundamental Plane relation \citep[FP;][]{Dressler1987, djorgovskidavis1987} tightly relates $R_{e}$, $\sigma$ and $\Sigma$ for bulge-dominated galaxies.

Galaxy scaling relations reflect physical mechanisms at work within galaxies. They enable us to gain deeper understanding of galaxy structure, formation and evolution. For example, \cite{Kassin2012} examined $V_{rot}/\sigma$ across redshift and found that galaxies accrete baryons faster earlier in their life-cycles, but that as galaxies evolve their accretion rate and gas content decrease; \cite{Obreschkow2014} demonstrated that the scaling relation between baryon angular momentum $(j)$, stellar mass $(M_*)$ and bulge fraction $(\beta)$ of spiral galaxies \citep[]{peebles69, Fall1983}, can produce and explain the FP (and FJ) scaling relation; \cite{Lagos2017}, using cosmological simulations, later confirmed the correlation between galaxy mass and specific angular momentum, and the evolution of the $M_*$--$j$ scaling relation in passive and active galaxies. Kinematic scaling relations are also useful in the study of the dark matter content of galaxies. For example, \cite{Desmond2017} used the FP to predict the amount of dark matter in the central regions of elliptical galaxies and suggested the deviation of the FP from the virial prediction (also known as the tilt of the FP) can be explained by non-homology in galaxy structure and variations in mass-to-light ratios; \cite{Ouellette2017} also found that the tilt of the FP correlates with the dark matter fraction of each galaxy. 

The TF relation applies to disk-dominated galaxies \citep[]{Bloom2017} while the FJ and FP relations apply to spheroidal galaxies. Incorporating galaxies of other morphologies into these scaling relations not only increases the scatter, but also changes the slopes and intercepts of the relations \citep[e.g.][]{Neistein1999, Iodice2003, Williams2010, Tonini2014}, consequently reducing the accuracy and reliability of the quantities derived. 

The scatter around the stellar mass versions of the FJ and TF relations can be reduced by replacing the rotation velocity or velocity dispersion, respectively, with the $S_{K}$ parameter introduced by \citet[]{Weiner2006}: 
\begin{equation}
	S_K=\sqrt{KV_{rot}^2+\sigma^2}
	\label{eq:SK}
\end{equation}
where $K$ is a constant, commonly taken to be 0.5 \citetext{e.g. \citealp{Weiner2006}, \citealp{Kassin2007}, \citealp{covington2010}, \citealp{Kassin2012}, \citealp{Cortese2014} \citepalias[hereafter:][]{Cortese2014}, \citealp{Simons2015}, \citealp{Straatman2017}, \citealp{Aquino-Ortiz2018} \citepalias[hereafter:][]{Aquino-Ortiz2018}}. By combining $V_{rot}$ and $\sigma$,  $S_{K}$ provides a common scaling relation for both early-type galaxies and late-type galaxies \citep[]{Kassin2007}. Furthermore, \citetalias{Cortese2014}, using data from the Sydney-AAO-Multi-object IFS (Integral Field Spectroscopy) Galaxy Survey \citep[hereafter SAMI Survey;][]{SAMIcroom2012, Bryant2015} early data release, and \citetalias{Aquino-Ortiz2018}, using data from the Calar Alto Legacy Integral Field Area survey \citep[CALIFA;][]{sanchez2012}, showed that $S_{K}$ can bring together the gas and stellar kinematic measurements of galaxies of all morphologies onto a single dynamical scaling relation. Numerical simulation has shown  $S_{K}$  is minimally affected by the  blurring effect due to seeing \citep{covington2010}. Therefore, $S_{K}$ is promising in the construction of a unified galaxy scaling relation that is robust with respect to morphologies and sub-optimal observing conditions. 

While $S_{K}$ has been a popular kinematic estimator and mass proxy, there are still a number of outstanding issues: (i)~while $K$ is commonly taken to be 0.5, this value has not been quantitatively justified; (ii)~\citetalias{Cortese2014} found a non-linearity in the $\log M_*$--$\log S_{K}$ relation below a stellar mass of $\sim10^{10}M_{\odot}$, but the existence and location of this change in slope (which determines the limit of validity for the scaling relation in low-mass galaxies) has not been reliably determined; and (iii)~there remains the question of how $S_{K}$ relates to aperture velocity dispersion ($\sigma_{\rm ap}$) from single-fibre surveys.

We use the latest data release from the SAMI Survey to expand on the work of \citetalias{Cortese2014}, and explore various aspects of  the $\log M_*$--$\log S_{K}$ scaling relation. Our work is structured as follows. In Section \ref{sec:datamethod}, we describe the data reduction, kinematic measurements, and sample morphologies. In Section \ref{sec:samiscalingrelation}, we construct $S_{K}$ from the gas and stellar measurements of our sample; compare $S_{K}$ to aperture velocity dispersion ($\sigma_{3^{\prime\prime}}$) measurements; and explore the sensitivity of the scatter of the relation for different values of $K$. In Section \ref{sec:discussion}, we compare our results to observations in the literature and discuss factors that influence the value of $K$. In Section \ref{sec:conclusion} we summarise our findings. We assume throughout a $\Lambda$CDM cosmology with $\Omega_M=0.3,\;\Omega_{\lambda}=0.7$ and $H_{0}=70$ km/s/Mpc. 

\begin{table*}
 \begin{tabular}{lllll}
  \hline\hline
  (1)    & (2)                                           & (3)                      & (4)                   & (5)                             \\
  Sample & Selection criteria                            & Number of galaxies with  & All have both gas and  & Used in                         \\
         & for each galaxy                               & gas/stellar measurements & stellar measurements? & Figure(s)                       \\ \hline
  A      & All of $V_{rot},\;\sigma,\;S_{0.5}$ must have & 410/270                  & False                 & \ref{fig:scalingrelations}     \\ 
         & less than 5\% error for each galaxy           &                          &                       &                                 \\ \hline
  B      & $S_{0.5}$ has error                           & 1256/1574                & False                 & \ref{fig:masshistogramdetailed} \\ 
         & less than 5\%                                 &                          &                       & \ref{fig:S05relations}         \\ \hline
  B1      & $\Delta V_i <20$ km/s, $\Delta \sigma_i<0.1\sigma+25$ km/s                           & 859/839                & False                 & \ref{fig:S05relations_clean} \\ 
         &  $S_{0.5}$ has error less than 5\%           &                          &                       &          \\ \hline
  C      & Measurements for gas                          & 223/223 ($V_{rot}$)      & True                  & \ref{fig:plotgasvsstars}        \\  
         & and stellar kinematics                        & 961/961 ($\sigma$)     & True                  &                                 \\ 
         & have error less than 5\%                      & 904/904 ($S_{0.5}$)      & True                  &                                 \\ \hline
  D      & Both $S_{0.5}$ and $\sigma_{3^{\prime\prime}}$& 864/1141                 & False                 & \ref{fig:S05_APER_SIG}          \\  
         & have error less than 5\%                      &                          &                       & \ref{fig:SAMISDSSTF}            \\ \hline
  E      & Both $V_{rot}$ and $\sigma$ for both          & 410/410 (ETG)            & True                  & \ref{fig:kvar}                 \\  
         & gas and stellar measurements                  & 232/232 (LTG)            & True                  &                                 \\ 
         & have error less than 10\%                     & 737/737 (All types)      & True                  &                                 \\ \hline
 \end{tabular}
 \caption{Sample selection criteria and description. All samples had the additional criterion of minor-to-major axis ratio less than 0.95. }
 \label{table:samplegroups}
\end{table*}

\section{Data and methods}
\label{sec:datamethod}

The SAMI Survey uses the AAOmega dual-beam spectrograph on the Anglo-Australian Telescope at Siding Spring Observatory \citep{sharp2006}. SAMI obtains integral field spectra by using 13 fused-fibre hexabundles, each containing 61 fibres \citep{Bland-Hawthorn2011,Bryant2014}. The SAMI spectra cover the wavelength range 3750--5750\,\AA\ at a resolution of $R$\,$\approx$\,1800, and  6300--7400\,\AA\ at a resolution of $R$\,$\approx$\,4300 \citep[]{Scott2018}. These give dispersion resolutions $\sigma_{\rm res}$ of 70\,km/s in the blue arm where we obtain the stellar kinematics, and 30\,km/s in the red for gas kinematics. 

The SAMI Survey includes both field and cluster galaxies \citep{Owers2017} with redshifts $0.004 < z < 0.095$, $r$-band Petrosian magnitudes $r_{\rm pet} <19.4$, and stellar masses $10^7$--$10^{12}M_{\odot}$. The stellar masses of SAMI galaxies are estimated as \citep[]{Bryant2015}:
\begin{align*}
	\log(M_*/M_{\odot}) =& -0.4i+0.4D-\log{(1+z)} \\
	                      &+(1.2117-0.5893z) \\
	                      &+(0.7106-0.1467z)\times (g-i)
	\label{eq:taylor}
	\numberthis{}
\end{align*}
where $M_*$ is the stellar mass in solar mass units, $D$ is the distance modulus, $i$ is the rest frame $i$-band apparent magnitude, and $g-i$ is the rest-frame colour of the galaxy, corrected for Milky-Way extinction \citep[]{Bryant2015}. More on the SAMI Survey and instrument can be found in \citet[]{SAMIcroom2012}.

\begin{figure}
	\includegraphics{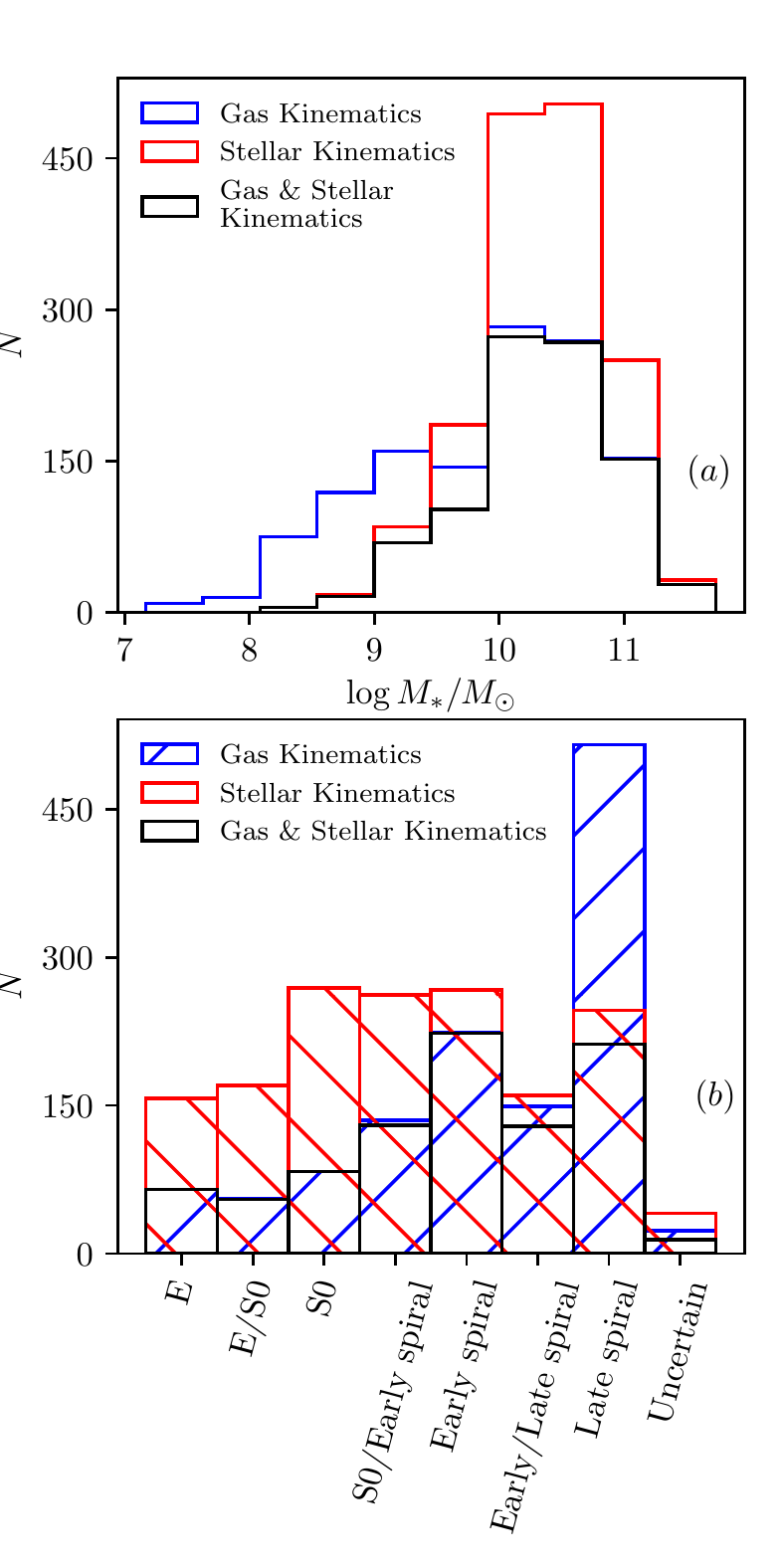}
	\caption{Panel~(a) shows the mass distributions of sample~B with various kinematic measurements available: red represents galaxies with stellar kinematics, blue represents galaxies with gas kinematics, and black represents galaxies with both gas and stellar kinematics. Panel~(b) shows the visual morphology distribution of sample~B: red bars represent galaxies with stellar kinematics, blue bars represent galaxies with gas kinematics, and black bars represent galaxies with both gas and stellar kinematics.}
    \label{fig:masshistogramdetailed}
\end{figure}

\subsection{Data Reduction and Sample}
\label{subsec:data}

SAMI data reduction consists of two stages, reducing raw data to row stacked spectra (RSS) using {\it 2dfdr}\footnote{https://www.aao.gov.au/science/software/2dfdr} and data cube construction from the RSS using the {\it SAMI Python} package \citep{allen2014}. The details of the data reduction and data cubing processes can be found in \citet[]{Allen2015}, \citet{sharp2015}, and \citet{Scott2018}.

We used gas and stellar kinematic maps extracted from SAMI internal data release v0.10 data cubes. For the stars, the velocity and velocity dispersion maps are measured using the penalised Pixel Fitting method \citep[pPXF;][]{pPXF}. pPXF extracts the stellar line-of-sight velocity distribution (LOVSD) in each spatial pixel (spaxel) from the observed spaxel spectrum assuming a Gaussian form:
\begin{equation}
	\mathcal{L}(v)=\frac{e^{-y^{2}/2}}{\sigma\sqrt{2\pi}}
	\label{eq:gausshermite}
\end{equation}
where $y = (v  -V )/ \sigma$. The $(V, \sigma )$ parameters of this model can be retrieved using a maximum likelihood optimisation. More details of the fitting routine can be found in \citet{vandesande2017}. At the time of this writing, the stellar kinematic data sample includes 2720 galaxies, all of which have also been fitted by the LZIFU \citep{ho2014} routine for $H\alpha$ emission line detection and measurement of the velocity and dispersion of the gas component using the one-component fit results.

Using the extracted SAMI stellar and gas kinematic maps, we select the spaxels as follows. First, spaxels are collected within an elliptical aperture with a semi-major axis of one effective radius. For all SAMI galaxies, their semi-major axis, position angles and ellipticity are determined using Multi-Gaussian Expansion \citep[MGE,][]{emsellem1994} fitting to $r$-band images from either the VLT Survey Telescope (VST) ATLAS \citep{Shanks2015, Owers2017} survey or the Sloan Digital Sky Survey (SDSS). Contrary to \citetalias{Cortese2014} and \citetalias{Aquino-Ortiz2018}, where spaxels are selected based on the absolute errors in velocity and velocity dispersion, we do not perform spaxel-level quality cuts other than requiring galaxies to have more than 5 spaxels within the aperture while rejecting empty spaxels. Instead, we perform an overall relative error cut on the kinematic parameter being investigated, only keeping galaxies with relative kinematic error less than 5--10\%. Rotation velocity, velocity dispersion, $S_{K}$ and associated error calculations are described in the next section. However, for comparison, we also produced scaling relations with similar quality criteria at the spaxel level, using only spaxels with velocity error $\Delta V_i <20$ km/s, velocity dispersion error $\Delta \sigma_i < 0.1 \sigma + 25$ \citep[]{vandesande2017b}, on top of the 5\% error cut on the final kinematic quantities. Details are shown in section \ref{subsec: spaxelclean}.

Depending on the kinematic parameter to be studied and the selection criteria, our parent sample of 2720 galaxies is divided into 5 sample groups (groups A--E). The selection criteria and sample group descriptions are listed in Table~\ref{table:samplegroups}. For investigation of the $\log{M_*}$--$\log{S_{K}}$ scaling relation we used sample~B. Sample~B includes gas-kinematics measurements for 1256 galaxies, stellar-kinematics measurements for 1574 galaxies, and 904 galaxies have both gas and stellar measurements. Galaxies in sample~B have a median of $\sim$70~spaxels. The stellar mass histogram in Figure~\ref{fig:masshistogramdetailed}a shows that sample~B is relatively complete in the high-mass ($\geq$ $10^{10}M_{\odot}$) range, but with sufficient numbers of galaxies in the low-mass range to constrain a scaling relation. 

\begin{figure*}
	\centering \includegraphics{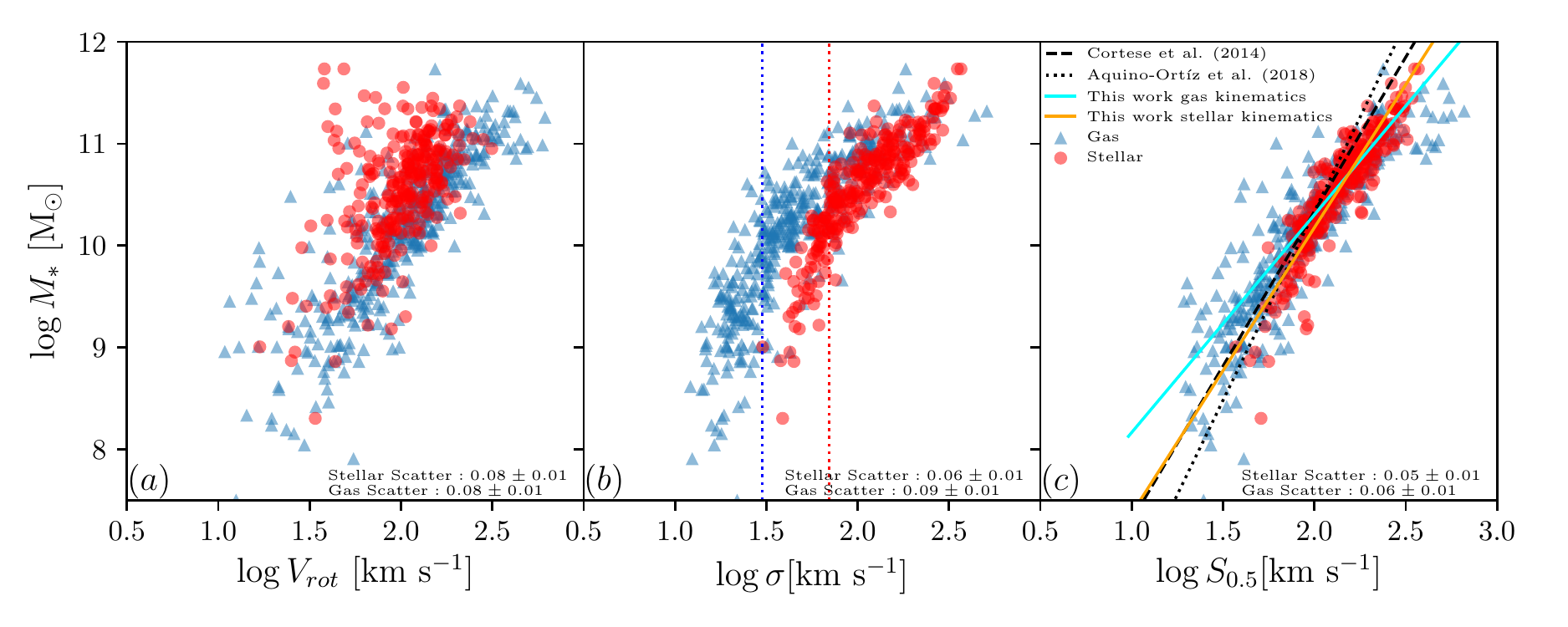}
	\caption{SAMI scaling relations from sample~A: (a)~Tully-Fisher, (b)~Faber-Jackson, and (c)~generalised $S_{0.5}$ scaling relation. Red dots represent galaxies with stellar measurements, blue triangles represent gas measurements. The observed scatter measured from the median absolute deviation for each scaling relation is annotated in each plot. In panel~(b), the red and blue vertical dotted lines represent SAMI spectral resolutions, $70$ km/s and $30$ km/s, respectively. In panel~(c) the orange solid line is the best fit line to the stellar $\log M_*$--$\log S_{0.5}$ scaling relation, and the cyan solid line is the best fit line to the gas scaling relation. Relations found by \citetalias{Cortese2014} and \citetalias{Aquino-Ortiz2018} are included for comparison; they are represented by the black dashed line and black dotted line respectively.}
    \label{fig:scalingrelations}
\end{figure*}

\subsection{Galaxy Kinematics}
\label{subsec:kinematics}

To calculate the rotation velocities of the gas and the stars, we use the velocity histogram technique, following \citet{Catinella2005} and \citetalias{Cortese2014}. The histogram technique is simple to implement and, in the process of calculating the velocity width, the Hubble velocity and peculiar velocity of the system naturally cancel. We calculate the velocity width $W$ between the 90th and 10th percentile points of the histogram of spaxel velocities within one $r$-band effective radius $(r_{e})$ elliptical aperture. Then we perform redshift $(z)$ and inclination $(i)$ corrections using an inclination angle estimated from the $r$-band minor-to-major axis ratio ($b/a$) which is obtained from the MGE fit to VST and SDSS images. We do not perform luminosity weighting on the rotation velocity. The rotation velocity is calculated as:
\begin{equation}
	V_{rot} = \frac{W}{2\;\sin{i}\;(1+z)}
	\label{eq:vrot}
\end{equation}
where the inclination of a galaxy is calculated as
\begin{equation}
	\cos{(i)} = \sqrt{\frac{(b/a)^2-q^2_{0}}{1-q^2_{0}}}
	\label{eq:inclination}
\end{equation}
and $q_{0}$ is the intrinsic axis ratio assumed to be 0.2. \citep[]{Catinella2012}. For edge-on galaxies with axis ratio less than 0.2, we do not perform inclination correction. We removed $\sim$60 galaxies that had axis ratio more than 0.95. We chose to retain edge-on as well as near face-on galaxies in our sample, because one of the main purposes of the study is to find a relation that is as inclusive as possible, without introducing a significant amount of outliers. We explored different sample constraints and found that $b/a>0.95$ was a reasonable compromise, excluding galaxies lying significantly far from the scaling relation, while being as inclusive as possible.

The effective velocity dispersion $\sigma$ of a galaxy is measured as the weighted mean of velocity dispersion measurements of each spaxel within an aperture radius of one effective radius, where the weight is the mean continuum flux: 
\begin{equation}
	\sigma^2 \equiv \frac{\sum_{i}L_{i}\sigma^2_{i}}{\sum_{i}L_{i}}
	\label{eq:sig}
\end{equation}
We highlight that we do not perform any spaxel-level quality cut here, other than having at least 5 non-empty spaxels. We then calculate $S_{0.5}$ as per Equation~\ref{eq:SK}, with $K=0.5$. We use bootstrapping to calculate the standard deviations of $V_{rot}$, $\sigma$ and $S_{0.5}$ to use as uncertainties. The bootstrap method involves randomly sampling the same number of spaxels as the total number of spaxels within the aperture, allowing for repeated selection of spaxels, and calculating $V_{rot}$, $\sigma$ and $S_{0.5}$. This step is carried out 1000 times to ensure the random samples represent the parent samples.

\subsection{Galaxy Morphologies}
\label{subsec:morphologies}

Galaxy morphologies in the SAMI sample vary from elliptical galaxies to late-type spiral and irregular galaxies. All SAMI galaxies are visually classified using the SDSS DR9 \citep[]{Ahn2012} RGB images by 12 members of the SAMI team following the classification scheme adopted in \citet{Kelvin2014}. Here we briefly summarise the classification scheme. First, judging by the presence/absence of a disk or spiral arms, the classifier determines whether a target is an early-type or late-type galaxy. Then in each class, classifiers visually determine if the galaxy contains a bulge (for late-type galaxies) or a disk (for early-type galaxies). Early-type galaxies with only a bulge, and without any disk component, are identified as elliptical (E) galaxies; early-type galaxies that show both bulge and disk components are identified as lenticular (S0) galaxies. Late-type galaxies all have spiral arms by classification definition; if there is a prominent bulge, they are classified as early-spiral galaxies; if there are only spiral arms without a central bulge, then they are classified as late-spiral or irregular galaxies. Where the SDSS image does not show enough features, or a consensus ($>$67\%) among classifiers is lacking, the galaxy is classified as Uncertain \citep{cortese2016}. For sample~B, where we have the most galaxies for studying the $\log{M_*}$--$\log{S_{0.5}}$ relation, their morphology distribution is shown in Figure~\ref{fig:masshistogramdetailed}b. There are relatively more early-type galaxies (E to S0 classes) in the stellar sample than the gas sample, and more late-type galaxies (early to late spiral classes) in the gas sample than the stellar sample.

Interacting galaxies, such as mergers, are typically removed from galaxy scaling relation studies. However, in our scaling relations, in order to obtain a scaling relation with minimal sample selection, we do not remove interacting galaxies from the main sample (sample B). The impact of merger galaxies is quantified further in section \ref{subsec:linearity}.

\begin{figure*}
	\centering 
	\hspace*{-1cm}\includegraphics[width=1.15\textwidth]{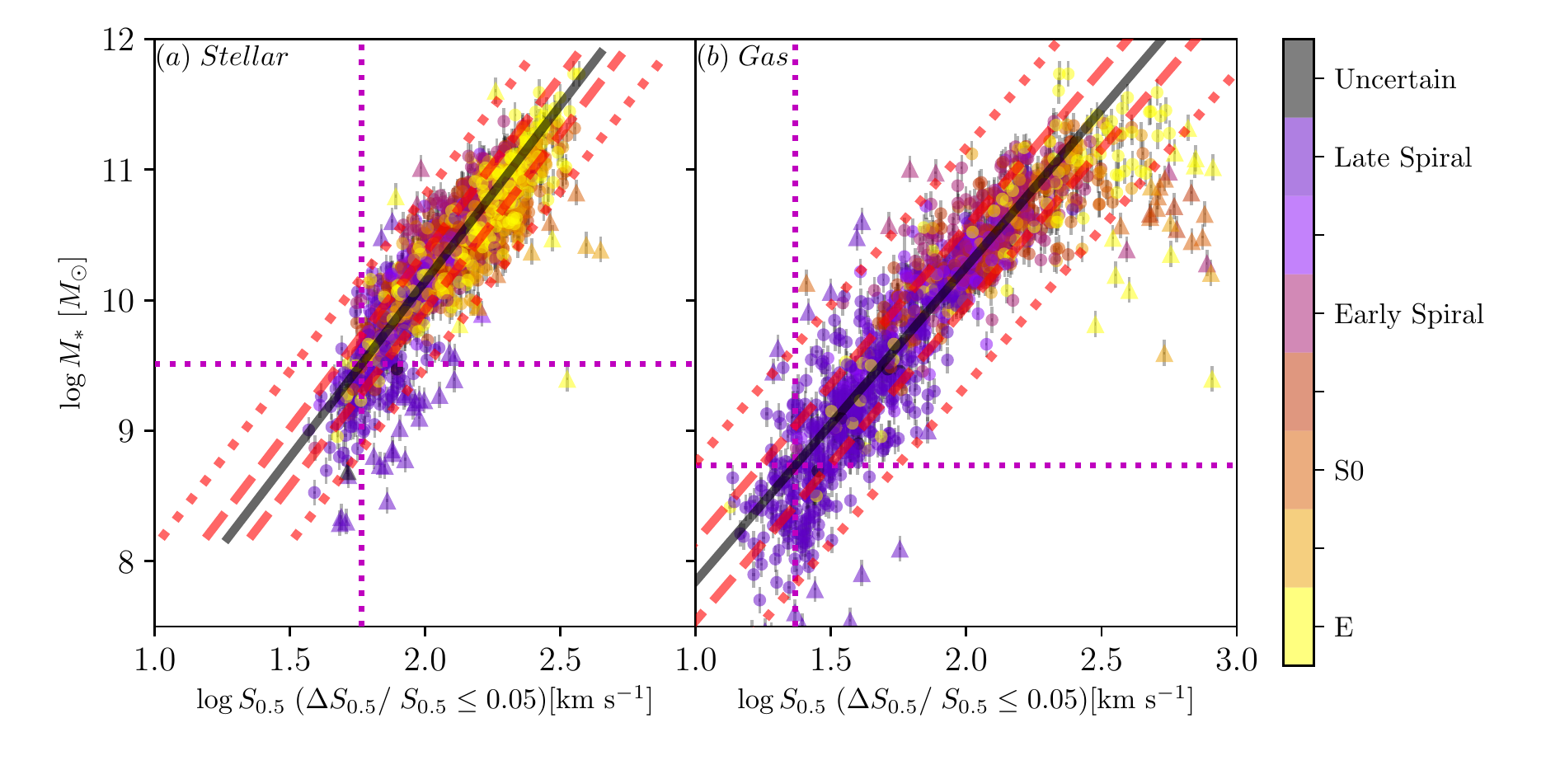}
	\caption{SAMI stellar and gas $S_{0.5}$ scaling relations from sample~B. The black solid line shows the line of best fit, with fitting parameters shown in Table \ref{table:fittingresult}. Red dashed and dotted lines show 1~and 3~RMS distance from the line of best fit. Triangular points are galaxies >3~RMS away from the line of best fit, and are excluded from the fitting routine. The magenta vertical and horizontal dotted lines show the location where the distribution of points deviate from a linear relation, which we fit as the sample limit. These limits occur at different $S_{0.5}$ values and different stellar masses for the stellar and gas samples: $S_{0.5,{\rm lim, stellar}} = 59$\,km/s for the stellar sample, $S_{0.5,{\rm lim, gas}} = 23$\,km/s for the gas sample.}
    \label{fig:S05relations}
\end{figure*}
\begin{figure*}
	\hspace*{-1cm}\includegraphics[width=1.15\textwidth]{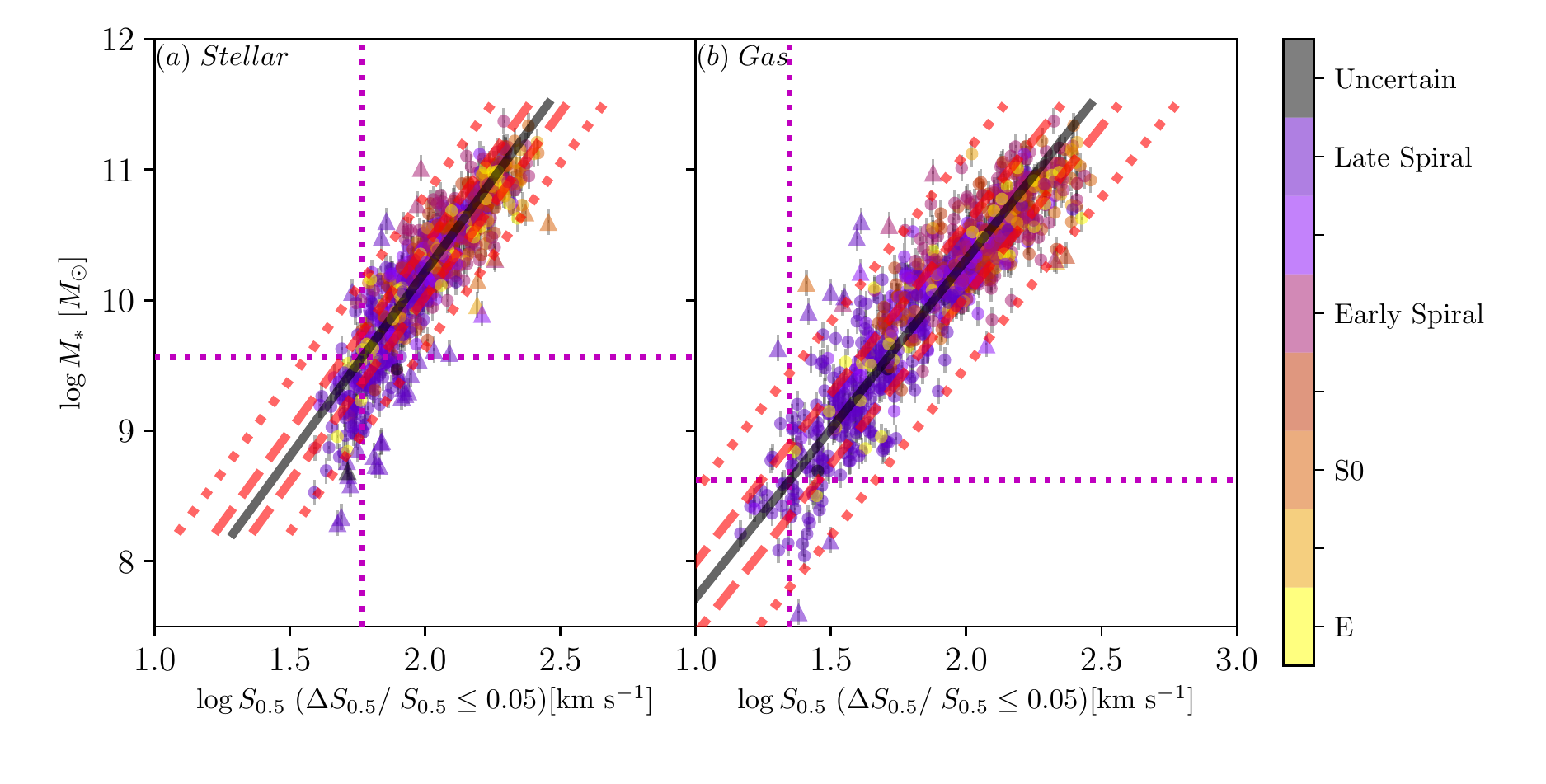}
	\caption{SAMI stellar and gas $S_{0.5}$ scaling relations from sample~B1. As for Figure \ref{fig:S05relations}, 
	except that the galaxy kinematics are measured after additional spaxel-level quality cuts as described in Table \ref{table:samplegroups}; fits are given in Table \ref{table:fittingresult}.}
    \label{fig:S05relations_clean}
\end{figure*}

\begin{table*}
 \begin{tabular}{lllccccc}
  \hline\hline
    Sample & Figure & $\rm Y$ & Slope (a) & Intercept (b) & $ Y_{\rm lim}$ [km/s] & $ M_{*, \rm lim}[M_{\odot}]$ & Scatter ($\rm MAD_{orth}$) \\ \hline 
    B       & 3a        & $S_{0.5, \rm stellar}$ & $0.37 \pm 0.01$ & $-1.77 \pm 0.06$ & $58.3\pm 1.0$ & $10^{9.56}$ & 0.048 $\pm$ 0.002 \\
            & 3b        & $S_{0.5, \rm gas}$     & $0.42 \pm 0.01$ & $-2.26 \pm 0.06$ & $23.0\pm 1.0$ & $10^{8.62}$ & 0.070 $\pm$ 0.002 \\ \hline 
    B1       & 4a        & $S_{0.5, \rm stellar}$ & $0.35 \pm 0.01$ & $-1.60 \pm 0.10$ & $58.3\pm 1.0$ & $10^{9.62}$ & 0.041 $\pm$ 0.002 \\
            & 4b        & $S_{0.5, \rm gas}$     & $0.39 \pm 0.01$ & $-1.99 \pm 0.05$ & $22.4\pm 1.0$ & $10^{8.56}$ & 0.063 $\pm$ 0.003 \\ \hline 
    D       & 7a        & $S_{0.5, \rm stellar}$ & $0.36 \pm 0.01$ & $-1.59 \pm 0.06$ & $67.0\pm 1.2$ & $10^{9.49}$ & 0.049 $\pm$ 0.001 \\
            & 7c        & $S_{0.5, \rm gas}$     & $0.40 \pm 0.01$ & $-2.16 \pm 0.06$ & $22.4\pm 1.0$ & $10^{8.73}$ & 0.061 $\pm$ 0.002 \\ \hline 
    D       & 7b        & $\sigma_{\rm stellar}$ & $0.38 \pm 0.01$ & $-1.82 \pm 0.06$ & $67.4\pm 1.1$ & $10^{9.61}$ & 0.068 $\pm$ 0.002 \\
            & 7d        & $\sigma_{\rm gas}$     & $0.48 \pm 0.01$ & $-2.98 \pm 0.14$ & $26.3\pm 1.1$ & $10^{9.17}$ & 0.090 $\pm$ 0.004 \\ \hline 
 \end{tabular}
 \caption{Scaling relation fitting results from \emph{hyper-fit}. All scaling relations have the form as described in Equation \ref{eq:models}.}
 \label{table:fittingresult}
\end{table*}

\section{SAMI SCALING RELATIONS}
\label{sec:samiscalingrelation}

\subsection{\texorpdfstring{$S_{0.5}$}{Lg} reduces scatter}

In this section we demonstrate the advantage of using the $S_{0.5}$ parameter in dynamical scaling relations in contrast to using $V_{rot}$ and $\sigma$ alone when using IFS data. In both \citetalias{Cortese2014} and \citetalias{Aquino-Ortiz2018}, the  $\log{M_*}$--$\log{S_{0.5}}$ scaling relation showed significant reduction in scatter when compared to the TF relation using $V_{rot}$ and FJ relation using $\sigma$. For comparison, we perform the same comparison between $S_{0.5}$, $V_{rot}$ and $\sigma$ using sample~A (as described in Table~\ref{table:samplegroups}). In sample~A, for each of the gas and stellar kinematic measurements, all of $S_{0.5}$, $V_{rot}$ and $\sigma$ have less than 5\% error for all galaxies. Figure~\ref{fig:scalingrelations} shows the correlation of stellar mass ($M_*$) with $V_{rot}$, $\sigma$, $S_{0.5}$ (i.e.\ the stellar mass TF, FJ, and combined $S_{0.5}$ scaling relations) as constructed from sample~A data. We perform maximum likelihood linear fitting to all the scaling relations in Figure~\ref{fig:masshistogramdetailed}, and measure their orthogonal median absolute deviations as their scatter. The fitting method is described in more detail in section~\ref{subsub:linefitting}; in this case we fit a simple linear relation with no cut-off. As can be seen from the annotated scatter values in the figure, for both the gas and stellar versions of these scaling relations, the $\log{M_*} - \log{S_{0.5}}$ relation consistently has less scatter than the TF and FJ relations. We also use morphologically-selected samples to compare the TF relation using LTGs, the FJ relation using ETGs, and the $\log M_* - \log S_{0.5}$ scaling relation using both LTGs and ETGs. $S_{0.5}$ continued to provide the tightest scaling relation. 

A caveat here is that our $V_{rot}$ measurements for late-type spiral galaxies do not reach the peak of their rotation curves, hence they cannot accurately trace the potentials of galaxies, and so, in our `TF' relation, $V_{rot}$ is not as good an estimator of $M_*$ as $S_{0.5}$.

The fits from \citetalias{Cortese2014} and \citetalias{Aquino-Ortiz2018} (from orthogonal fitting of the combined gas and stellar mixed sample) are shown in Figure~\ref{fig:scalingrelations}c by the dashed line and dotted line respectively. There are small differences between the slopes for our gas and stellar samples, shown in orange and cyan respectively, both with each other and with the linear relations found by \citetalias{Cortese2014} and \citetalias{Aquino-Ortiz2018}. However, given the differences in sample selection, survey systematics, and fitting methods, it is hard to interpret the observed differences in slope as physical differences. 

\subsection{Linearity of the \texorpdfstring{$S_{0.5}$}{Lg} scaling relation}
\label{subsec:linearity}

\citetalias{Cortese2014} constructed the gas FJ and $\log M_*$--$\log S_{0.5}$ scaling relations and observed that the slope became steeper for low-mass ($M_*<10^{10}M_{\odot}$) galaxies. This change in slope is also present in our FJ relation in Figure~\ref{fig:scalingrelations}b, and the $\log M_*$--$\log S_{0.5}$ relations in Figure~\ref{fig:scalingrelations}c. A bend in these kinematic scaling relations is expected, as the fitted linear relations would otherwise predict zero motions for low-mass galaxies ($\sim10^{5}-10^{6}M_{\odot}$). The cause of the bend will be discussed in more detail in Section~\ref{subsec:discussion1}, but for now it is crucial to take the bend into consideration in fitting the scaling relation. To locate the change in slope in our $\log M_*$--$\log S_{0.5}$ scaling relations more precisely, we investigate the scaling relations in detail with sample~B in Figure~\ref{fig:S05relations}, where galaxies are selected only based on $\Delta S_{0.5}/S_{0.5} \leq 0.05$.

\subsubsection{Straight line with a knee}
\label{subsub:linefitting}

To find the point at which the slope of the relation changes, we hypothesise that there exists a sample limit at position $(S_{0.5,\rm lim}, M_{*,\rm lim})$ where a single linear model can no longer describe the distribution of the sample. For all stellar mass measurements below this $M_{*,\rm lim}$ value $S_{0.5}$ values will be normally distributed around a limiting value $S_{0.5,\rm lim}$. For stellar masses above the $M_{*,\rm lim}$ value the scaling relation is assumed to be a linear relation described by:
    \begin{equation}
        \log{S_{0.5}} = a\log{M_*} + b
        \label{eq:scalingrelation}
    \end{equation}
We then use this combination of a linear model with a constant limit cut-off in our maximum likelihood fitting routine, assuming $\log{M_{*,i}}$ and $\log{S_{0.5,i}}$ for each galaxy have Gaussian uncertainties $\sigma_{\log{M_{*,i}}}$ and $\sigma_{\log {S_{0.5,i}}}$ respectively. The total posterior logarithmic likelihood $\ln \mathcal{L}$ under this model is 
\begin{equation}
	\ln\mathcal{L}=\frac{1}{2}\sum_i\left[\ln\frac{a^{2}+1}{s_{\log S_{0.5,i }}^{2}}-\frac{\left(\log S_{0.5,i }-Y\right)^{2}}{s_{\log S_{0.5,i }}^{2}}\right]
	\label{eq:log_likelihood}
\end{equation}
where $Y$ is a linear function above $M_{*,\rm lim}$ and a constant below $M_{*,\rm lim}$, namely,
\begin{equation}
    Y = 
    \begin{cases} 
         a\log{M_{*, i}} +b, & M_*>M_{*,\rm lim}\\ 
         a\log{M_{*,\rm lim}}+b, & M_*\leq M_{*,\rm lim}\\
    \end{cases} 
    \label{eq:models}
\end{equation}
and $s_{\log S_{0.5,i}}^{2}\equiv\sigma_{\log S_{0.5}}^{2}+\sigma_{M_{*,i}}^{2}a^{2}+\sigma_{\log S_{0.5,i}}^{2}$ where $\sigma_{\log S_{0.5}}$ is the intrinsic scatter of about the model. By adjusting the fitting parameters $a,\;b,\;\sigma_{\log S_{0.5}},\;\log M_{*,\rm lim}$ and using Markov Chain Monte Carlo \citep[MCMC;][]{emcee2013} we can find the model parameters that maximise the likelihood given by Equation~\ref{eq:log_likelihood} together with their uncertainties. To ensure the model is robust against outliers, we repeat the fitting routine five times whilst rejecting points that are $>3\sigma$ away each time (represented by triangular points in Figure \ref{fig:S05relations}). The fitting method is described in more detail as the 2D {\em hyper-fit} in \citet[]{RobothamObreschkow2015}.

\subsubsection{The bends in the scaling relations}
\label{subsub:bendyfitting}

Following the fitting method described in the previous section, we fit the linear+cutoff model to our sample~B galaxies, as shown in Figure~\ref{fig:S05relations}; the fitted parameters and their uncertainties are given in Table \ref{table:fittingresult}. 

There are 40-50 outliers in our stellar and gas scaling relations, mostly at $\log S_{0.5} \geq 2.5$. In the stellar scaling relation, visual inspection of these galaxies shows them to be contaminated by either foreground stars or (in clusters) nearby bright galaxies. In the gas scaling relations, these galaxies are generally ETGs with relatively larger errors in their gas kinematic measurements. 13 of the outliers in the gas scaling relation are found to be merger galaxies. Removing these merger galaxies did not change the slope, intercept or the scatter of the scaling relation. All of the outliers disappear when we apply a more stringent quality cut than $\Delta S_{0.5}/S_{0.5} \leq 0.05$.

The fitted $M_{*,\rm lim}$ for each of the gas and stellar versions of the scaling relation can be converted to $S_{0.5,\rm lim}$ using the model. For the stellar version of the scaling relation, the bend occurs at $({M_{*, \rm lim, stellar}},\;{S_{0.5, \rm lim, stellar}}) =(10^{9.6}\,M_{\odot}, 61\,{\rm km/s})$, and for the gas version, $({M_{*,  \rm lim, gas}},\;{S_{0.5,  \rm lim, gas}}) =(10^{8.9}\,M_{\odot}, 23\,{\rm km/s})$. The fact that the bend in the stellar and gas scaling relations occurs at different stellar mass values suggests the nature of the bend in our scaling relation is unlikely to be a physical phenomenon. 

\begin{figure*}
	\hspace*{-1cm}\includegraphics[width=1.1\textwidth]{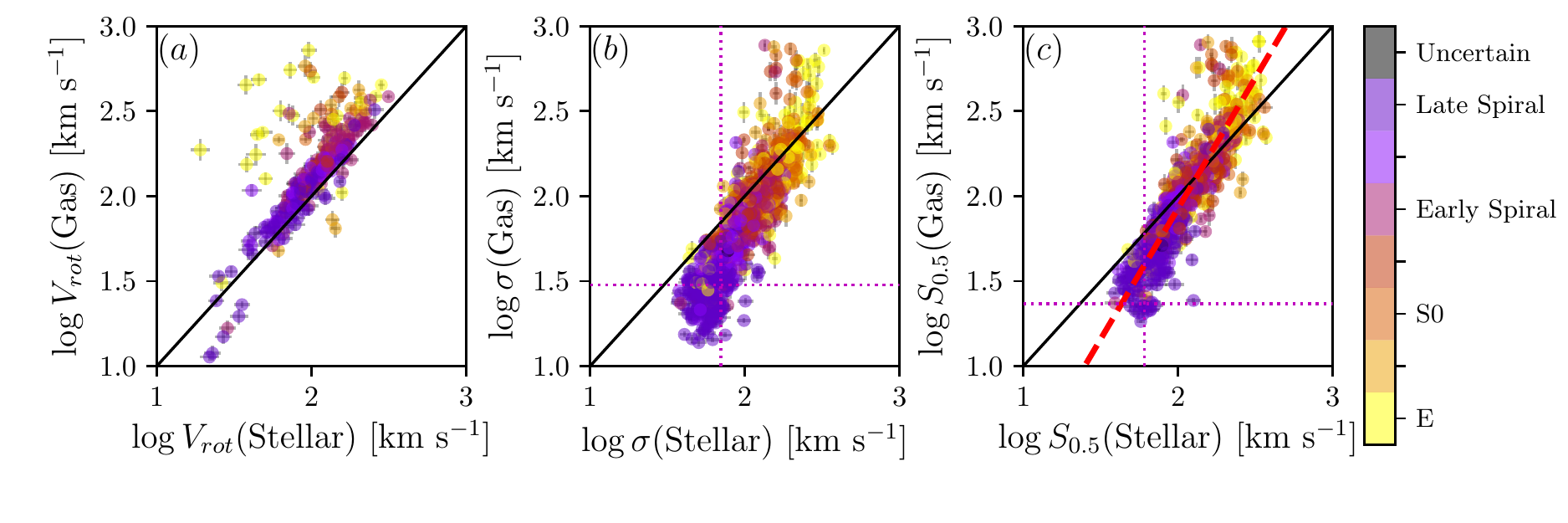}
	\caption{Comparison between SAMI sample~C gas and stellar measurements of (a)~$V_{rot}$, (b)~$\sigma$, and (c)~$S_{0.5}$. Galaxies are colour-coded by morphology. For visibility, we show 3 standard deviation error bars. Black solid lines in each panel show the one-to-one relation. The red dashed line in panel~(c) is the best-fit line to the points. Horizontal and vertical magenta lines in panel~(b) show the gas and stellar instrumental dispersions (30\,km/s and 70\,km/s respectively) and in panel~(c) show the fitted gas and stellar sample limits (23\,km/s and 59\,km/s respectively).}
    \label{fig:plotgasvsstars}
\end{figure*}

\subsection{Effect of spaxel-level quality cut}
\label{subsec: spaxelclean}

The $\log{M_{*}}-\log{S_{0.5}}$ scaling relation in our study with sample B is constructed with minimal sample selection as well as minimal quality cut on the spaxels, in order to be as inclusive as possible. This procedure is different from \citetalias{Cortese2014} and \citetalias{Aquino-Ortiz2018}. To investigate the importance of high quality spaxels, we perform a similar spaxel-level cut where we only keep spaxels with velocity error $\Delta V <20$ km/s, velocity dispersion error $\Delta \sigma < 0.1 \sigma + 25$, on top of the 5\% error cut on the $S_{0.5}$ parameter. The velocity dispersion spaxel selection routine follows \cite{vandesande2017b}, which ensures spaxels have $S/N >3 \text{\AA}^{-1}$ for $\sigma>35$ km/s.
These criteria produce sample B1. Figure \ref{fig:S05relations_clean} shows the scaling relations produced with sample B1, and Table \ref{table:fittingresult} shows the fitting results. 

By introducing spaxel-level quality cuts, the slopes of both stellar and gas scaling relation decreased by 0.02 to 0.03 (i.e. by 2 to 3$\sigma$). The scatters of both scaling relations were also reduced significantly. The locations of the sample limits ($Y_{\rm lim}$, $M_{*,\rm lim}$) remained the same. It is clear that performing spaxel-level quality selection can increase the quality of the scaling relations. However, by applying a spaxel-level quality cut, sample sizes were reduced by 30\% to 50\%. For our study, it is better to have larger sample sizes than scaling relations with less scatter, as the sample will be divided further in later sections.

\begin{figure*}
	\includegraphics[width=1.1\textwidth]{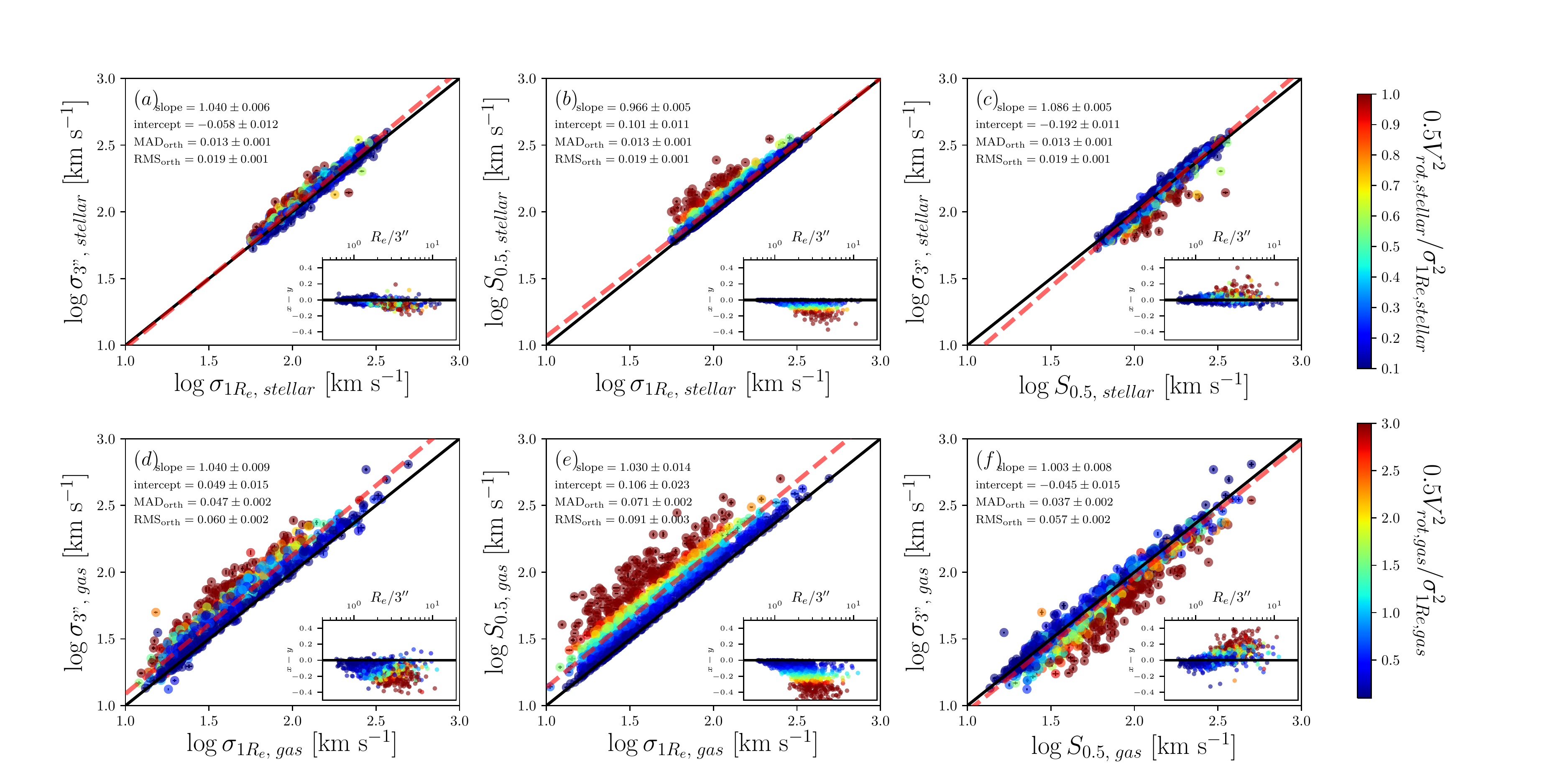}
	\caption{Direct comparison, using sample~D, between $S_{0.5}$ and $\sigma_{3^{\prime\prime}}$ (the 3-arcsecond-diameter aperture velocity dispersion), and average velocity dispersion $\sigma_{1R_e}$ within $1R_e$ from gas and stellar kinematics. Their differences against apparent galaxy size are plotted in the inset plots. Black solid lines are the one-to-one relations; red dashed lines are the best fits; points are colour coded by $0.5V^2_{rot}/\sigma^2$ to indicate the dominant term in the $S_{0.5}$ parameter.}
    \label{fig:S05_APER_SIG}
\end{figure*}

\begin{figure*}
	\includegraphics{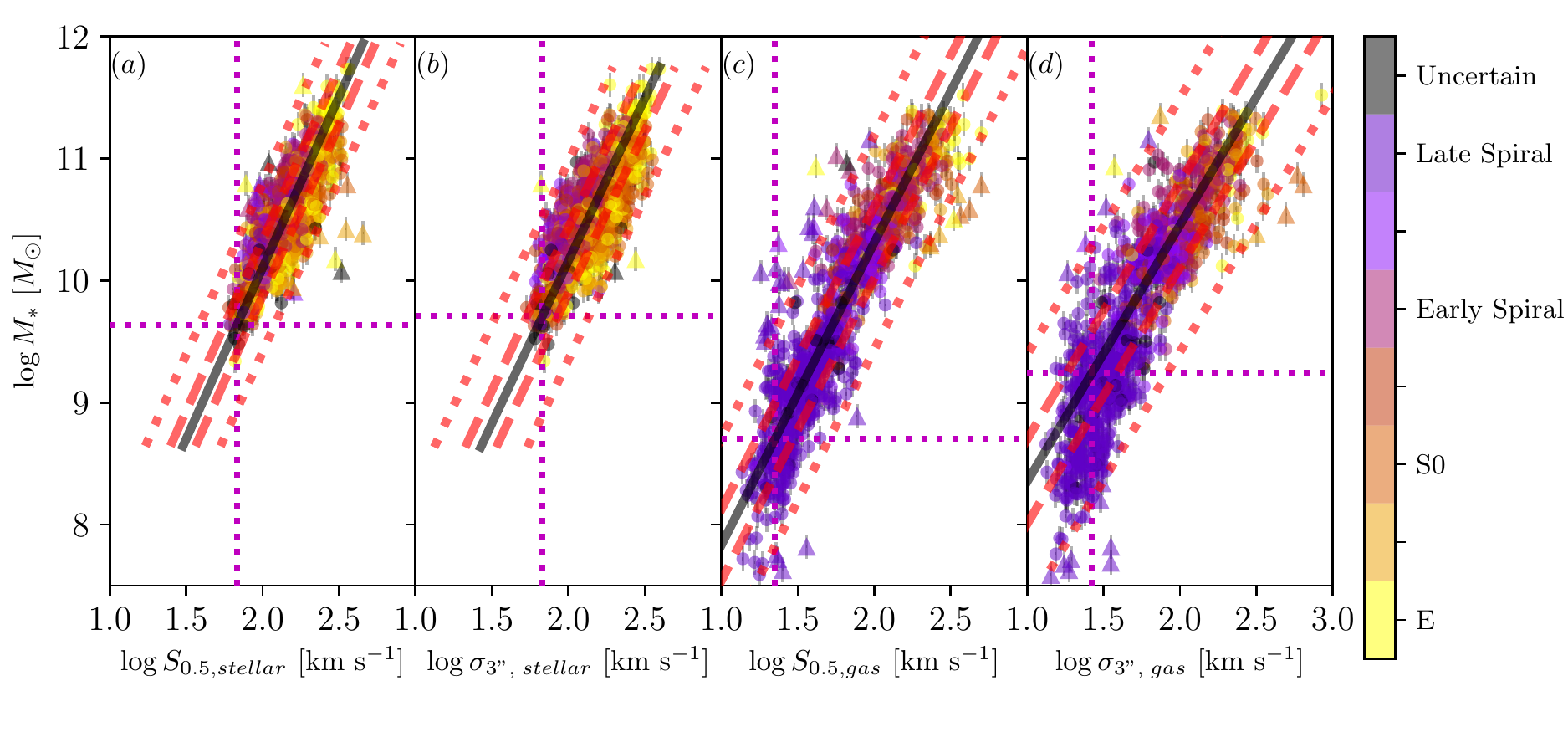}
	\caption{Scaling relations constructed from $S_{0.5}$ kinematic parameters and aperture velocity dispersions using sample~D. Panels~(a), (b), (c) and~(d) show scaling relations constructed between stellar mass and, respectively, $S_{0.5, stellar}$, $\sigma_{3^{\prime\prime}, stellar}$, $S_{0.5, gas}$ and $\sigma_{3^{\prime\prime}, gas}$ measurements. Black solid lines in each panel are lines of best fit; red dashed and dotted lines define distances 1\;RMS and 3\;RMS away from the line of best fit. Triangular points are measurements $>$3\;RMS away from the line of best fit, and are excluded from the fit as outliers. Horizontal and vertical magenta dotted lines are the fitted model limits. Fit parameters and uncertainties are given in Table~\ref{table:fittingresult}. For both gas and stellar measurements, $S_{0.5}$ consistently produces scaling relations with less scatter than aperture velocity dispersion.}
    \label{fig:SAMISDSSTF}
\end{figure*}

\subsection{Gas and stellar \texorpdfstring{$S_{0.5}$}{Lg} disagreement}
\label{subsec:gasvsstars}

To test whether the $S_{0.5}$ parameters from the stellar and gas kinematics trace the same gravitational potentials, we compare the rotation velocities, velocity dispersions and $S_{0.5}$ measurements of stellar and gas components on a per-galaxy basis with SAMI samples~C (as described in Table~\ref{table:samplegroups}). Galaxies in the sample are selected to have both gas and stellar kinematic errors less than 5\% for each of $V_{rot}, \sigma$ and $S_{0.5}$. Figure~\ref{fig:plotgasvsstars}a, using sample~C, shows that stars in general rotate more slowly than the gas. This is due to asymmetric drift, where the rotation velocity of the stars is lower than that of the gas because stars have additional pressure support against gravity from a higher dispersion \citep[]{BinneyTremaine08}. The mean ratio between the stellar and gas rotation velocities is $\langle V_{rot,stellar}/V_{rot, gas}\rangle \sim 0.77\pm0.29$ which is consistent with the value ($\sim$0.75) obtained by \citetalias{Cortese2014}. 

Using sample~C, where both gas and stellar $\sigma$ measurements have less than 5\% error, the offset between the SAMI gas and stellar FJ relations observed in \citetalias{Cortese2014} is reflected in our FJ relation. The mean ratio between gas and stellar dispersions is $\langle \sigma_{stellar}/\sigma_{gas} \rangle \sim1.58\pm0.59$, as shown in Figure \ref{fig:plotgasvsstars}b, consistent with the value ($\sim$1.55) found by \citetalias{Cortese2014}. We note that $\sigma_{stellar}>\sigma_{gas}$ for galaxies with $\log{\sigma_{gas}} <2.5$ is expected due to asymmetric drift. The galaxies with $\log{\sigma_{gas}} >2.5$ that lie significantly above the one-to-one line are almost entirely elliptical and S0 galaxies; these galaxies have low gas abundance, making it harder to obtain accurate $\sigma_{gas}$ measurements and resulting in relatively larger uncertainties in $\sigma_{gas}$, as shown in Figure~\ref{fig:plotgasvsstars}b.

In \citetalias{Cortese2014}, the gas and stellar $S_{0.5}$ parameters are found to have a mean logarithmic difference (gas\,$-$\,stellar) of $-$0.02\,dex. In our sample~C, the logarithmic difference is $-$0.05\,dex. In Figure~\ref{fig:plotgasvsstars}c, while the one-to-one line (black) goes through our $S_{0.5, gas}-S_{0.5, stellar}$ distribution, the best-fit (red-dashed) line has a slope of $1.49 \pm 0.02$, which suggests a systematic disagreement between $S_{0.5, gas}$ and $S_{0.5, stellar}$. One explanation of $S_{0.5, stellar}>S_{0.5, gas}$ for $S_{0.5}<100$\,km/s can be traced to the fact that $\sigma_{stellar}>\sigma_{gas}$. However, inclusion of $V_{rot, gas}$ component where $V_{rot, gas}>V_{rot, stellar}$ does not seem to sufficiently compensate $S_{0.5 gas}$ to bring balance between $S_{0.5, stellar}$ and $ S_{0.5, gas}$. The scatter in the $S_{0.5, gas}-S_{0.5, stellar}$ correlation increases for galaxies with $\log S_{0.5, gas}>2.25$. As in Figure~\ref{fig:plotgasvsstars}b, this increase in scatter can be explained by E and S0 galaxies having larger uncertainties in their $\sigma_{gas}$ measurements. However, in the {\em hyper-fit} routine, measurements are error-weighted, reducing the impact of points with large errors; moreover, with a tighter uncertainty cut (using galaxies with $<$2\% errors) the slope remained significantly above unity ($1.34 \pm 0.04$). Lastly, restricting the sample to only include galaxies with both gas and stellar measurements above the fitted sample limit, the slope remained steeper than unity at $1.49\pm 0.02$. This disagreement between stellar and gas $S_{0.5}$ parameters requires investigation on a per-galaxy basis, especially for lower mass ($\log M_*< 9.5$) galaxies where $S_{0.5, stellar} > S_{0.5, gas}$. This will be studied in more detail in future.

\subsection{IFS and aperture kinematic measurements}
\label{subsec:samivssdss}

While IFS data provides resolved spatial information on galaxy kinematics, compared to single fibre observations it is observationally expensive. 
In order to compare the effect of
IFS data on the scaling relations, we constructed aperture spectra from SAMI data cubes within a 3-arcsecond-diameter (SDSS-like) aperture. Applying a 5\% error quality cut to the aperture velocity dispersions, we obtained $\sigma_{3^{\prime\prime}, gas}$ for 864 galaxies, and $\sigma_{3^{\prime\prime}, stellar}$ for 1141 galaxies; these form sample~D (see Table~\ref{table:samplegroups}). Figure~\ref{fig:S05_APER_SIG} shows a comparison between average velocity dispersion $\sigma_{1R_e}$, $S_{0.5}$ parameter and $\sigma_{3^{\prime\prime}}$, and their logarithmic difference as a function of $R_e/3^{\prime\prime}$, the galaxy size relative to the aperture diameter. Galaxies are colour-coded by their $0.5V^2_{rot}/\sigma^2$ ratio to show the balance between $V_{rot}$ and $\sigma$ within the $S_{0.5}$ parameter, galaxies with $0.5V^2_{rot}/\sigma^2$ ratio above and below the colour bar limits are shown in solid red and solid blue respectively.

For stellar measurements, the kinematics is mostly pressure-supported and dominated by $\sigma_{1R_e}$, which is confirmed by 1098 out of 1141 galaxies having $0.5V^2_{rot}/\sigma^2<1$. The inset residual plots comparing the residuals to galaxy size indicate that galaxies with $0.5V^2_{rot}/\sigma^2>1$ have $R_e > 6 ^{\prime\prime}$, and they are furthest away from the one-to-one line in all panels. From panel a, the increasing $0.5V^2_{rot}/\sigma^2$ ratio from the line-of-best-fit confirms that $\sigma_{3^{\prime\prime}}$ contains a $V_{rot}$ component. Comparing panel a and c, we can see that galaxies with $0.5V^2_{rot}/\sigma^2>1$ are affected the most by replacing $\sigma_{1R_e}$ with $S_{0.5}$. Since these large galaxies with high $0.5V^2_{rot}/\sigma^2>1$ ratio make up less than 5\% of the sample, the scatter remained the same between $\sigma_{3^{\prime\prime}}-\sigma_{1R_e}$ and $\sigma_{3^{\prime\prime}}-S_{0.5}$ comparisons. 

The disparity due to the $V_{rot}$ component between $\sigma_{1R_e}$, $S_{0.5}$ and $\sigma_{3^{\prime\prime}}$ is amplified for the gas kinematics, shown in the lower panels of Figure \ref{fig:S05_APER_SIG}. Gas has relatively more rotation support than stars; 386 out of 864 galaxies have $0.5V^2_{rot}/\sigma^2>1$, which makes up 44\% of the sample (for stellar measurements, 4\%). The range of $0.5V^2_{rot}/\sigma^2$ values is larger compared to the stellar measurement, as shown by the values on the colour bar.

For both gas and stellar components, comparing $\sigma_{3^{\prime\prime}}$ to $\sigma_{1R_e}$ shows that $\sigma_{3^{\prime\prime}}$ contains additional rotation support. However, by replacing $\sigma_{1R_e}$ with $S_{0.5}$ (effectively adding a $0.5V^2_{rot}$ component), $S_{0.5}$ for large ($R_e > 6 ^{\prime\prime}$) galaxies became significantly larger than their $\sigma_{3^{\prime\prime}}$ values. This indicates that the rotation velocity component within $\sigma_{3^{\prime\prime}}$ measurements is weighted less than that within the $S_{0.5}$ parameter. It is clear that if one wishes to match $\sigma_{3^{\prime\prime}}$ and $S_{0.5}$, the $K$ value in the definition of $S$ must be less than 0.5, but greater than 0. For our particular purpose of constructing scaling relations, we compare $\log M_* - \log \sigma_{3^{\prime\prime}} $ and $\log M_* - \log S_{0.5}$ in the following section.

Even though the $S_{0.5}$ parameter and 3 arcsec aperture velocity dispersion $\sigma_{3^{\prime\prime}}$ for the gas and the stars cover different parts of galaxies, they remain in broad agreement. This agreement is encouraging because while the $\log M_* - \log S_{0.5}$ relation applies to galaxies of all morphologies, obtaining the $S_{0.5}$ parameter requires observationally-expensive IFS data, whereas the measuring the velocity dispersion in single-fibre surveys is observationally relatively cheap. The residual plots in Figure~\ref{fig:S05_APER_SIG}c indicates that $\sigma_{3^{\prime\prime}}$ is a fairly unbiased predictor of $S_{0.5,stellar}$ out to about $R_e=3^{\prime\prime}$ (rather than, as one might have expected, $R_e=3^{\prime\prime}/2$); moreover, the scatter only grows relatively gradually beyond $R_e=3^{\prime\prime}$, and more slowly for the stellar measurements than for the gas measurements.

\subsection{Comparing IFS and aperture scaling relations}
\label{subsec:FJrelations}

As both aperture $\sigma$ and $S_{0.5}$ are used in kinematic scaling relations, we compare variants of the FJ relation using $S_{0.5}$ and $\sigma_{3^{\prime\prime}}$ in Figure~\ref{fig:SAMISDSSTF}. For both stellar and gas versions, using the $S_{0.5}$ parameter consistently provides tighter relations with less scatter than using $\sigma_{3^{\prime\prime}}$ (see Table~\ref{table:fittingresult}). This confirms that $S_{0.5}$ is a better mass proxy than single-aperture velocity dispersion, and suggests the promising possibility of using $S_{0.5}$  to reduce the scatter in other scaling relations such as the FP relation \citep[e.g.][]{Graham2017}. On the other hand, for many purposes the slight increase in scatter in the scaling relation that results from using the aperture dispersion rather than $S_{0.5}$ ($\sim$\,0.02\,dex for stars and $\sim$\,0.03\,dex for gas) may be an acceptable trade-off for the lower observational cost of single-fibre surveys relative to IFS surveys.

We notice in Figure \ref{fig:SAMISDSSTF} that choosing $S_{0.5}$ over $\sigma_{3^{\prime\prime}}$ yields more outliers (triangular points, excluded from the fit) that are $>3$\,RMS (red dotted line) away from the line of best fit (black line). This is due to factors such as inclination errors and individual spaxel quality. IFS sampling radius affects the quality of the $S_{0.5}$ parameter more than single-aperture velocity dispersion measurements. In SAMI $\sigma_{3^{\prime\prime}}$ measurements, spectra from each spaxel are co-added to form the aperture spectrum, which increases the signal-to-noise ratio, and returns more reliable (albeit less accurate) kinematic measurements.

\subsection{Varying \texorpdfstring{$K$}{Lg}}
\label{subsec:Kvariation}

\begin{table}
 \begin{tabular}{lccc} \hline\hline
                     & Gas All     & Gas ETG     & Gas LTG     \\
    Optimal $K$      & 0.4         & 0.4         & 0.2         \\
    $\rm MAD_{orth}$ & 0.076       & 0.079       & 0.057       \\ \hline
                     & Stellar All & Stellar ETG & Stellar LTG \\
    Optimal $K$      & 0.7         & 0.3         & 0.2         \\
    $\rm MAD_{orth}$ & 0.045       & 0.044       & 0.047       \\ \hline 
 \end{tabular}
 \caption{Values of $K$ that return the minimum scatter for gas and stellar scaling relations, for each morphological sample.}
 \label{table:kvalues}
\end{table}

The original $S_{K}$ parameter introduced by \citet{Weiner2006} combines the galaxy rotation velocity and velocity dispersion in quadrature, weighting the rotation velocity by the factor $K$. This is commonly taken to be $K=0.5$, which is correct only for virialised systems with spherical symmetry and isotropic velocity dispersion \citep{Kassin2007}. We empirically test the effect of changing the value of $K$ in the construction of the $S_{K}$ parameter by measuring the scatter about the scaling relations. In this section we use sample~E, where each galaxy has both $V_{rot}$ and $\sigma$ with less than 10\% error for both gas and stellar kinematics. We chose 10\% error on $V_{rot}$ and $\sigma$ to provide a statistically large sample for determining the scatter about the scaling relation, though the results are qualitatively unaffected if we use a more stringent sample selection. 

We tested $K$ values ranging from 0 to 3 in the $\log M_*-\log S_{K}$ scaling relation, and measured the orthogonal median absolute deviation from the scaling relation for each $K$ value. We performed this test for both the gas and stellar versions of the scaling relation, and for all galaxies as well as separately for early-type (E and S0) and late-type (Sp and Irr) galaxies. We then measured the scatter in the scaling relation at every $K$ value for each of these samples.

\begin{figure}
\centering 
\includegraphics{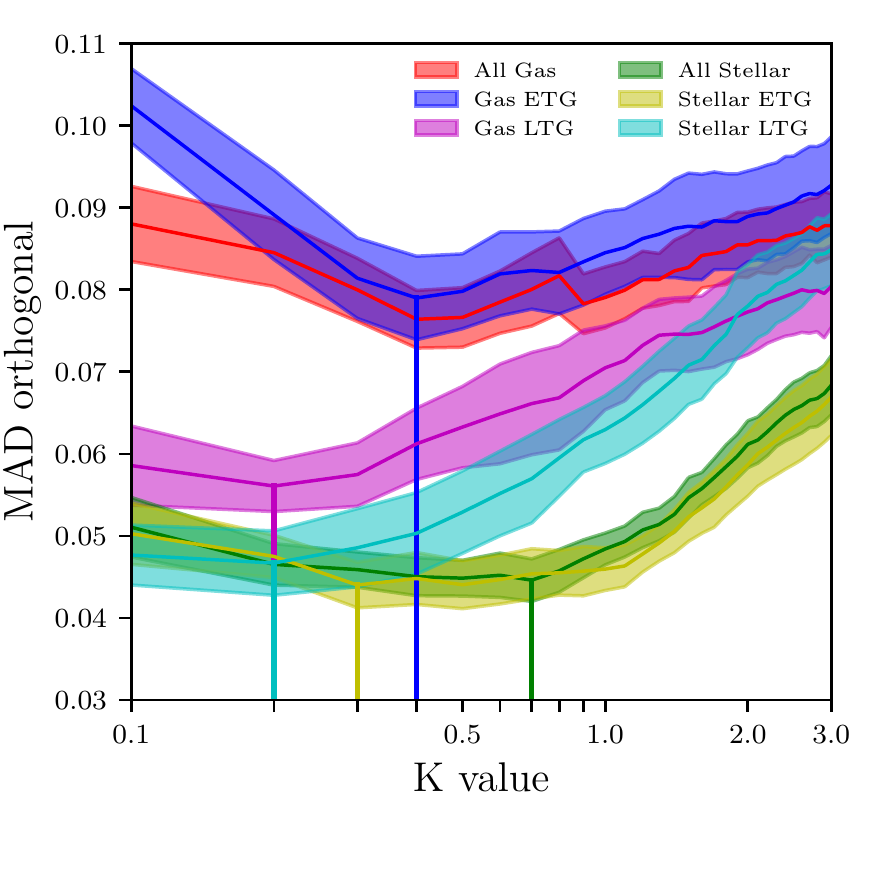}
\caption{Effect of the value of $K$ on the scatter of the SAMI gas and stellar $\log M_*-\log S_{K}$ scaling relations. The curves show the scatter about the relation for each colour-coded galaxy sample and the shaded regions show the $1\sigma$ uncertainties. The gas and stellar samples are further divided into ETGs and LTGs to determine the effect of morphology. Where $S_{K, gas}$ is used, the full, ETG and LTG samples are colour-coded red, blue and purple respectively; where $S_{K, stellar}$ is used, the full, ETG and LTG samples are colour-coded green, yellow and cyan respectively. The vertical lines show the locations of the minimum scatter for each sample (n.b. the red and blue vertical lines are over-plotted).}
\label{fig:kvar}
\end{figure}

Figure~\ref{fig:kvar} shows the effect of varying $K$ in the $\log M_*-\log S_{K}$ scaling relation for gas and stellar measurements, for both the full sample and for early-type and late-type galaxies; the values of $K$ that return the minimum scatter for each version of the scaling relation are listed in Table~\ref{table:kvalues}. For the full sample and for early-type galaxies the minimum in the scaling relation scatter is broad and spans approximately $K=0.3$--0.7. Given the uncertainties in the scatter measurements, shown as shaded regions in Figure~\ref{fig:kvar}, $K=0.5$ returns a scatter consistent with the minimum when either ETGs or all galaxies are considered. This consistency justifies the common usage of $K=0.5$ in the literature and the consistency of analyses when $K$ is varied \citetext{e.g. \citealp{covington2010},\citetalias{Cortese2014},\citetalias{Aquino-Ortiz2018}}. However, the late-type galaxy scaling relations, whether based on stellar or gas kinematics, have scatter that is minimised over a narrower range around $K=0.2$, and marginally higher scatter for $K=0.5$. While this small coefficient for $V_{rot}$ seems counter-intuitive given $V_{rot}$ is the primary component in the TF relation for LTGs, it is apparent that for LTGs their $\sigma$ component must be taken into consideration. The need for properly including the $\sigma$ component can be due to effects such as beam-smearing where $V_{rot}$ decreases and $\sigma$ increases. In the $S_{K}$ parameter, $\sigma$ is up-weighted by down-weighting the $V_{rot}$ component, so that by using the $S_{K}$ parameter, the effect of beam smearing is minimised. For a broad sample of the galaxy population, the optimum value depends on the mix of morphologies in the sample; the specific factors that lead to this situation are discussed in the following section. 

\section{discussion}
\label{sec:discussion}

\subsection{SAMI scaling relations}
\label{subsec:discussion1}

Using integral field spectroscopy (IFS) from the SAMI Survey for a parent sample of 2720 galaxies, we re-examine the $\log M_* - \log S_{0.5}$ scaling relation studied in \citetalias{Cortese2014}. We confirm that the $S_{0.5}$ kinematic parameter, measured from either the stars or the gas, brings galaxies of all morphologies onto a common scaling relation with stellar mass. The slopes of the scaling relation obtained here, both for the stars ($0.37\pm0.01$) and for the gas ($0.42\pm0.01$), are steeper than that ($0.34\pm0.01$) obtained by \citetalias{Cortese2014}. This difference is likely due to the difference in stellar mass distributions, as our sample is skewed towards higher masses ($M_* > 10^{10}\,M_{\odot}$); \citetalias{Cortese2014} in fact fitted for galaxies with $M_* > 10^{10}\,M_{\odot}$ and found a steeper slope. This difference in slope is slightly decreased when we performed spaxel-level quality cut which eliminated some of most massive galaxies ($M_{*}>10^{11.5}M_{\odot}$).

\citetalias{Cortese2014} also observed a change in slope across a mass range of $8.5 \lesssim \log M_*/M_{\odot} \lesssim 11.5$; this apparent change in slope is also present in our sample. \citetalias{Cortese2014} suggested this slope change could be due to not accounting for the mass in neutral gas; due to lack of HI data, we are unable to test this hypothesis. However, the change in slope in our $ \log M_* - \log S_{0.5}$ scaling relation could be due to other reasons besides not accounting for the total baryonic mass.  Possible causes include: a kinematic measurement limit due to the combined effects of the instrumental resolution and S/N ratio, as a reference we presented the SAMI instrument resolutions for gas and stellar kinematics in Figure~\ref{fig:scalingrelations}; the uncertainty in the low velocity dispersion measurements positively skewing the distribution of the observed velocity dispersions; a surface brightness limit causing the low-mass sample to be biased towards brighter galaxies with relatively high velocity dispersions; an intrinsic physical effect causing low-mass galaxies to have higher velocity dispersions than expected based on the linear relation for high-mass galaxies. Unfortunately, with the currently available SAMI data and its limitations in S/N ratio, spectral resolution and sample selection, we cannot distinguish all of these possible causes.

However, we fitted a linear scaling relation with a cutoff at a sample limit $M_{*,\rm lim}$ (corresponding to $S_{0.5, \rm lim})$ and found that the sample limits in the stellar and gas versions of the relation occur at different masses. This suggests the bend observed in our sample is unlikely to be caused by intrinsic properties such as stellar mass (this is not to say that there is no physical change in the slope of the scaling relation at some lower mass). \citetalias{Aquino-Ortiz2018} suggest galaxies with stellar mass below $\log M_*/M_{\odot} \sim 9.5$ have more dark matter content within the effective radius as the mass decreases, so that the dynamical mass (from $S_{0,5}$ estimation) to stellar mass ratio for low mass galaxies increases, resulting in a change in slope in $\log M_* - \log S_{0.5}$. Unfortunately, while we do have galaxies in our sample with $\log M_*/M_{\odot} < 9.5$, the range of low-mass galaxies does not extend down to $10^{7}$--$10^{8}\,M_{\odot}$. For future work, high S/N IFS observations of low-mass galaxies with higher spectral resolution ($\sigma_{\rm res} \sim 10$ km/s) will be necessary to fully determine the linearity of the stellar scaling relation throughout the $7 \lesssim \log M_*/M_{\odot} \lesssim 12$ mass range \citep[e.g. using Hector,][]{ByrantHector2016}.

\subsection{IFS and aperture kinematics}
\label{subsec:discussion2}

In Section \ref{subsec:samivssdss} and \ref{subsec:FJrelations} we compared measurements of the $S_{0.5}$ parameter and the aperture velocity dispersion and found surprisingly good agreement between the two kinematic tracers (as seen in Figure~\ref{fig:S05_APER_SIG}). This agreement is interesting because, while both $S_{0.5}$ and $\sigma_{3^{\prime\prime}}$ are measures of galaxy internal motions, they measure those motions differently. By definition, $\sigma_{3^{\prime\prime}}$ measures the second moment of the LOSVD integrated over a 3-arcsecond-diameter aperture, including the effect of rotation velocity. For $S_{0.5}$, the $\sigma$ component is a luminosity-averaged quantity from LOSVD dispersions measured locally over an aperture (here up to $1R_e$). By combining these local dispersion measurements with a global rotation measurement $V_{rot}$ via a suitable scaling factor, $S_{0.5}$ produces tighter scaling relations with $M_*$ than $\sigma_{3^{\prime\prime}}$ (as seen in Figure~\ref{fig:SAMISDSSTF} and Section~\ref{subsec:FJrelations}). 

The extra information provided by IFS and the more complex calculation involved in deriving $S_{0.5}$ thus provide a better understanding of this scaling relation (and others). However, IFS is observationally expensive while fibre surveys are observationally cheap. So for purposes requiring very large samples (e.g.\ exploring the effect of environment on scaling relations for galaxies of different morphological types or using scaling relations to derive distances and peculiar velocities) aperture dispersions may be a more efficient and economical choice.

\subsection{The importance of \texorpdfstring{$K$}{Lg}}
\label{subsec:discussion3}

The motivation for using $K=0.5$ in the $S_{K}$ parameter originates from the virial theorem prediction of the relation between circular velocity and velocity dispersion for an isothermal sphere, $V_{circ}\propto\sqrt{\alpha} \cdot \sigma$, where $\alpha$ is a constant that describes the density profile of the system. We have found empirically that the scatter depends weakly on the value of $K$, with minimum scatter occurring between $K=0.2$ and $K=0.7$. 

There are a number of possible factors that can theoretically influence the value of $K$: 

(1) {\it Solutions to the Jeans equation}. The convention of $K=0.5$ originates from the singular sphere case of the Jeans equation, where the circular velocity is given by:
\begin{equation}
    V^2_{circ}=\frac{GM(r)}{r}=-\sigma^2 \frac{d\ln \rho}{d \ln r} 
	\label{eq:vcirc}
\end{equation}
where $\rho$ and $r$ are the density and radius. For isothermal spheres, $\rho \propto \sigma^2 / r^2 $, and at large radius $d\ln \rho/d \ln r \sim -2$; therefore $V^2_{circ} \sim 2 \sigma^2$ \citep[][Section~4.3.3b]{BinneyTremaine08}. As we measure $V_{rot}$ via the velocity width technique, $V^2_{rot}\equiv V^2_{circ} \sim 2\sigma^2$, so $K=0.5$ corresponds to equally weighting $V_{rot}$ and $\sigma$, which would be optimal if they have similar uncertainties. Note that this conclusion makes assumptions about the galaxy density profile, the radius at which the kinematics are measured, and the relative precision of the $V_{rot}$ and $\sigma$ measurements---almost no real galaxies or kinematic observations satisfy all these assumptions. Nonetheless, as we have seen in Section~\ref{subsec:Kvariation}, $K=0.5$ is still close to optimal. 

(2) {\it Velocity distribution function}. The value of $K$ depends on the velocity distribution function of a galaxy, and in particular on the bulge-to-disk ratio and the $V/\sigma$ ratio for each of the bulge and disk components. In the case of pressure-supported systems with negligible rotation, the average stellar line-of-sight velocity dispersion $\overline{\sigma}_{LOS}$ is a weighted sum of directional components $\sigma_r$, $\sigma_\theta$ and $\sigma_\phi$. Excluding observational artefacts, the combination of components is dictated by the anisotropy parameter \citep[Eqn~4.61]{BinneyTremaine08}
\begin{equation}
	\beta \equiv 1- \frac{\sigma^2_{\theta}+\sigma^2_{\phi}}{2\sigma^2_{r}}
\label{eq:beta}
\end{equation}
Depending on whether the distribution function of stars is tangentially biased ($\beta<0$), radially biased ($\beta > 0$) or isotropic ($\beta=0$), the combination of $\sigma_{r}$, $\sigma_\theta$ and $\sigma_\phi$ making up $\overline{\sigma}_{LOS}$ will be different. Thus the $K$ value needs to be adjusted to correct for the unobserved components of $\overline{\sigma}_{LOS}$.

(3) {\it Observational\;artefacts}. Since the optimal $K$ value is determined by comparing the scatter in the $\log M_* - \log S_{0.5}$ relation, the quality of kinematic parameter measurements and the scatter of the scaling relation are crucial. \citet{covington2010} have shown with numerical simulations that instrument blurring effects such as spatial resolution and seeing, which contribute to the scatter in the TF relation, do not show significant effects on the measured $S_{0.5}$ values. \citetalias{Aquino-Ortiz2018} also performed a detailed kinematic analysis with spatially resolved rotation velocity measurements. They found that the $S_{0.5}$ parameter consistently reduced the scatter in scaling relations, taking into account the uncertainties in the $V_{rot}$ measurement for dispersion-dominated systems. In Section~\ref{subsec:discussion1}, we noted that there could be multiple extrinsic causes for a non-linear scaling relation and/or increased scatter, including S/N ratios, instrument resolution, sample selection, and kinematic uncertainties. Thus the best $K$ value is determined by a combination of intrinsic dynamical properties and observational artefacts. In order to use the $\log M_*- \log S_{K}$ scaling relation to predict physical attributes of observed systems, it is crucial to make sure the scatter in the scaling relation is not dominated by systematic error.  

\section{Conclusions}
\label{sec:conclusion}

In this paper we present the  $\log M_*- \log S_{0.5} $ scaling relation constructed from the SAMI Galaxy Survey. The $S_{0.5}$ parameter is useful in bringing galaxies of all morphologies onto a common relation. With no sample pruning (other than S/N quality cuts) the scatter in the $\log M_*- \log S_{0.5}$ relation is significantly less than the TF and FJ relations constructed from the same sample. Interestingly, applying only a relative error cut on $S_{0.5}$, without any spaxel-level quality cut, still provides a tight scaling relation. Both the stellar and gas versions of the $\log M_*- \log S_{0.5} $ scaling relation have a sample limit where the relation deviates from a linear relation. We found the sample limits occur at different stellar masses for the gas and stellar samples, implying that the apparent non-linearity in the relation is not physical; this is emphasised by the fact that the $S_{0.5, \rm lim }$ values corresponding to these mass cut-offs are proportional to the effective instrumental resolutions for the stellar and gas measurements. 

Comparing $S_{0.5}$ to single-aperture velocity dispersion $\sigma_{3^{\prime\prime}}$ shows excellent agreement between the two parameters. For the gas measurements the residuals $\sigma_{3^{\prime\prime}, gas}-S_{0.5, gas}$ trend negatively with galaxy angular size, while for the stellar measurements the residuals $\sigma_{3^{\prime\prime}, stellar}-S_{0.5, stellar}$ show no correlation with galaxy angular size. In constructing the mass scaling relations, $S_{0.5}$ consistently produced less scatter than $\sigma_{3^{\prime\prime}}$. 

In order to test the importance of choosing an optimal value of $K$ in the construction of the $S_{k}$ parameter, we measured the scatter of the scaling relations at different values of $K$. By investigating the correlation between the scatter of the scaling relation and the value of $K$ in the $S_{K}$ parameter, we found that for both stellar and gas measurements $K=0.5$ is close to producing the minimum scatter for samples containing only ETGs or mixtures of ETGs and LTGs; however, for samples containing only LTGs $K=0.2$ gave significantly less scatter.

These findings are broadly consistent with previous studies by \citetalias{Cortese2014} using early release of SAMI data and \citetalias{Aquino-Ortiz2018} using CALIFA data. 
 
The $S_{0.5}$ kinematic parameter allows the construction of a robust and inclusive galaxy scaling relation with relatively little scatter. The tight correlation between $S_{0.5}$ and $\sigma_{3^{\prime\prime}}$ implies that a scaling relation with only slightly greater scatter can be constructed for galaxies of all morphologies using large-scale single-fibre galaxy surveys.

\section*{Acknowledgements}
We thank the referee for their constructive report. The SAMI Galaxy Survey is based on observations made at the Anglo-Australian Telescope. The Sydney-AAO Multi-object Integral field spectrograph (SAMI) was developed jointly by the University of Sydney and the Australian Astronomical Observatory. The SAMI input catalogue is based on data taken from the Sloan Digital Sky Survey, the GAMA Survey and the VST ATLAS Survey. The SAMI Galaxy Survey is supported by the Australian Research Council Centre of Excellence for All Sky Astrophysics in 3 Dimensions (ASTRO 3D), through project number CE170100013, the Australian Research Council Centre of Excellence for All-sky Astrophysics (CAASTRO), through project number CE110001020, and other participating institutions. The SAMI Galaxy Survey website is \href{http://sami-survey.org}{sami-survey.org}.

DB is supported by an Australia Government Research Training Program Scholarship and ASTRO 3D. FDE acknowledges funding through the H2020 ERC Consolidator Grant 683184. JBH is supported by an ARC Laureate Fellowship that funds JvdS and an ARC Federation Fellowship that funded the SAMI prototype. JJB acknowledges support of an Australian Research Council Future Fellowship (FT180100231). JvdS is funded under Bland-Hawthorn's ARC Laureate Fellowship (FL140100278). NS acknowledges support of a University of Sydney Postdoctoral Research Fellowship. Parts of this research were conducted by ASTRO 3D, through project number CE170100013. LC is the recipient of an Australian Research Council Future Fellowship (FT180100066) funded by the Australian Government. SB acknowledges the funding support from the Australian Research Council through a Future Fellowship (FT140101166). SMC acknowledges the support of an Australian Research Council Future Fellowship (FT100100457). BG is the recipient of an Australian Research Council Future Fellowship (FT140101202). MSO acknowledges the funding support from the Australian Research Council through a Future Fellowship (FT140100255). Support for AMM is provided by NASA through Hubble Fellowship grant \#HST-HF2-51377 awarded by the Space Telescope Science Institute, which is operated by the Association of Universities for Research in Astronomy, Inc., for NASA, under contract NAS5-26555. SKY acknowledges support from the Korean National Research Foundation (2017R1A2A1A05001116) and by the Yonsei University Future Leading Research Initiative (2015-22-0064). This study was performed under the umbrella of the joint collaboration between Yonsei University Observatory and the Korean Astronomy and Space Science Institute. 



\bibliographystyle{mnras}
\bibliography{mybibfile}




\appendix



\bsp	
\label{lastpage}
\end{document}